\documentclass[11pt]{article}             

\usepackage{amsmath,amssymb,amsthm}
\usepackage{mathbbol,bbm}
\newcommand{\Cx}{\mathbb C}
\newcommand{\idty}{\Eins}
\DeclareMathOperator*{\tr}{Tr}
\DeclareMathOperator{\meo}{MEO}
\newcommand{\<}{\langle}
\renewcommand{\>}{\rangle}
\providecommand{\abs}[1]{\lvert#1\rvert}
\providecommand{\norm}[1]{\lVert#1\rVert}
\newcommand{\s}[1]{\mathsf{#1}}
\renewcommand{\r}[1]{\mathrm{#1}}
\allowdisplaybreaks[4]

\begin{document}

\begin{center}
\textbf{\large
 Note on multiple additivity of minimal Renyi entropy output of
 the Werner--Holevo channels} \\[6pt]
{\large R.~Alicki}\footnote{E-mail: fizra@univ.gda.pl} and
{\large M.~Fannes}\footnote{E-mail: mark.fannes@fys.kuleuven.ac.be} \\[6pt]
\emph{Institute for Theoretical Physics and Astrophysics, University of
Gda\'nsk, ul.\ Wita Stwosza 57, PL-80-952 Gda\'nsk, Poland \\
Instituut voor Theoretische Fysica, K.U.Leuven, Celestijnenlaan~200D,
B-3001 Heverlee, Belgium}
\end{center}
\bigskip

\noindent
\textbf{Abstract:}
We give an elementary self-contained proof that the minimal entropy output
of arbitrary products of channels $\rho \mapsto \frac{1}{d-1}\bigl(
\idty - \rho^{\s T} \bigr)$ is additive.
\medskip

This paper is concerned with efficiency of transmission of classical
information through a particular quantum channel. Generally, the
minimal entropy output of a general quantum channel $\Gamma$, given
in terms of a completely positive trace-preserving map on the
$d$-dimensional density matrices,  quantifies the
minimal noise produced by the channel in terms of the von~Neumann entropy of
transformed pure states
\begin{equation*}
 \meo\bigl[ \Gamma \bigr] := \inf_{\varphi,\ \norm{\varphi}=1} \r S\Bigl(
 \Gamma\bigl(|\varphi\>\<\varphi|\bigr) \Bigr)
\end{equation*}
where $\r S(\rho)$ is the von~Neumann entropy of a density matrix $\rho$.
Recalling the definition of the $p$-Renyi entropy of a density matrix
$\rho$, $p>1$
\begin{equation*}
 \r S_p(\rho) := - \frac{1}{p-1} \log\Bigl(\tr \rho^p \Bigr),
\end{equation*}
the minimal $p$-Renyi entropy output is
\begin{equation}
 \meo_p\bigl[ \Gamma \bigr] := \inf_{\varphi,\ \norm{\varphi}=1}
 \r S_p\bigl(\Gamma(|\varphi\>\<\varphi|)\bigr).
\label{meop}
\end{equation}
Obviously, $\lim_{p\downarrow1} \meo_p\bigl[\Gamma\bigr] =
\meo\bigl[\Gamma\bigr]$ and we can therefore extend the notation $\meo_p$
to $p=1$.

The $p$-Renyi entropy of a density matrix is directly related to its
$p$-norm. Let $X$ be a general, not necessarily square, complex matrix. The
$p$-norm of $X$ is given by
\begin{equation*}
 \norm{X}_p := \Bigl( \tr \abs{X}^p\Bigr)^{\frac{1}{p}} = \Bigl( \sum_j \chi_j^r
 \Bigr)^{\frac{1}{p}}
\end{equation*}
where $\abs{X} = \sqrt{X^*X}$ and where the $\chi_j$ are the (non-zero)
singular values of $X$. The $p$-norm is the non-commutative analogon of the
usual $p$-norm of sequences of complex numbers. As a density matrix is
positive semi-definite, we have the immediate relation
\begin{equation}
 \r S_p(\rho) = - \frac{p}{p-1} \log \norm{\rho}_p.
\label{ren-pnorm}
\end{equation}
In terms of $p$-norms, the quantity corresponding to $\meo_p(\Gamma)$ is
\begin{equation*}
 \nu_p(\Gamma) := \sup_{\varphi,\ \norm{\varphi}=1} \Bigl\Vert
 \Gamma\bigl( |\varphi\>\<\varphi| \bigr) \Bigr\Vert_p.
\end{equation*}
Using~(\ref{ren-pnorm}) and~(\ref{meop}) we find
\begin{equation}
 \meo_p(\Gamma) = - \frac{p}{p-1} \log \nu_p(\Gamma).
\label{meop-nup}
\end{equation}

The additivity conjecture
\begin{equation}
 \meo\bigl[\Gamma_1\otimes\Gamma_2\bigr] =
 \meo\bigl[\Gamma_1\bigr] + \meo\bigl[\Gamma_2\bigr]
\label{conmeop}
\end{equation}
means that entanglement
cannot help to lower the noise in parallel channels, described by tensor
products. Recently, the
relevance of this conjecture has been stressed, as it was shown to be
equivalent to other important additivity issues in quantum information
theory, such as Holevo capacity and entanglement of formation,
see~\cite{sho}. There is a similar conjecture for $\meo_p$, $1<p\le2$.
As both the von~Neumann and the Renyi entropies are additive
on tensor products, the conjecture amounts to showing that
\begin{equation}
 \meo_p\bigl[ \Gamma_1\otimes\Gamma_2 \bigr] \ge \meo_p\bigl[ \Gamma_1 \bigr] +
 \meo_p\bigl[ \Gamma_2 \bigr].
\label{conmeop2}
\end{equation}
The relation~(\ref{meop-nup}) allows us to restate the
conjecture~(\ref{conmeop}) as multiplicativity of maximal $p$-norms
\begin{equation*}
 \nu_p\bigl[ \Gamma_1\otimes\Gamma_2 \bigr] = \nu_p\bigl[ \Gamma_1 \bigr]
 \nu_p\bigl[ \Gamma_2 \bigr].
\end{equation*}

We shall consider here a particular quantum channel, known as Werner--Holevo
channel
\begin{equation}
 \Gamma_d(\rho) = \frac{1}{d-1}\bigl(\idty - \rho^{\s T} \bigr).
\label{cha}
\end{equation}
Here, $A^{\s T}$ denotes the transpose of $A$ w.r.t.\ the standard basis of
$\Cx^d$. The map $\Gamma_d$ has both a high symmetry and a truly
entangling nature compared e.g.\ with the depolarising channel. In fact,
$U \Gamma_d(\rho)U^* = \Gamma_d(\overline U\rho(\overline U)^*)$ where $U$
is an element of SU(d) and $\overline U$ is its complex conjugate and it is
the non-trivial extreme point in the set of all such maps. In the case
$d=3$, it is even an extreme points in the set of all completely positive
trace-preserving maps. This channel has
often been considered in the literature, see e.g.\ \cite{hol,wer,dat}.
We show that the minimal entropy output of $\otimes^N \Gamma_{d_j}$ is
multiply additive for $1\le p\le2$
\begin{equation}
 \meo_p\bigl[\otimes_{j=1}^N \Gamma_{d_j}\bigr] = \sum_{j=1}^N
 \meo_p\bigl[\Gamma_{d_j}\bigr].
\label{add}
\end{equation}
In the particular case $p=1$, $N=2$ and $d_1=d_2$,
formula~(\ref{add}) has been proven in~\cite{dat} using more advanced
techniques, see also~\cite{mat}.
Because of~(\ref{meop-nup}) this is equivalent to showing the
multiple multiplicativity of the maximal $p$-norms, see~\cite{kin}
\begin{equation*}
 \nu_p\bigl[\otimes_{j=1}^N \Gamma_{d_j}\bigr] = \prod_{j=1}^N
 \nu_p\bigl[\Gamma_{d_j}\bigr],\quad 1<p\le2.
\end{equation*}

A straightforward computation yields
\begin{align*}
 \meo_p\bigl[ \Gamma_d \bigr]
 &:= \inf_{\varphi,\ \norm{\varphi}=1} \r S_p\Bigl(
 \Gamma_d\bigl(|\varphi\>\<\varphi|\bigr) \Bigr) \\
 &=\inf_{\varphi,\ \norm{\varphi}=1} \r S_p\Bigl({\textstyle \frac{1}{d-1}} \bigl( \idty -
 |\overline\varphi\>\<\overline\varphi| \bigr)\Bigr) = \log(d-1).
\end{align*}
Here, $\overline\varphi$ is the complex conjugate of $\varphi$ in the
standard basis.

Next, we write out the action of the channel $\otimes_{j=1}^N
\Gamma_{d_j}$ on the pure state generated by a normalised vector
$\Omega\in\otimes_{j=1}^N \Cx^{d_j}$
\begin{equation*}
 X_N(\Omega) := \otimes_{j=1}^N \Gamma_{d_j}\bigl( |\Omega\>\<\Omega| \bigr) =
 \prod_{j=1}^N \frac{1}{d_j-1} \sum_{\Lambda\subset\{1,2,\ldots, N\}}
 (-1)^{\abs{\Lambda}}\, \rho_\Lambda \otimes \idty_{\Lambda^{\r C}}.
\end{equation*}
Here, $\rho_\Lambda$ is the reduced density matrix of
$|\overline\Omega\>\<\overline\Omega|$ to the sites in $\Lambda$.
We now compute
\begin{align}
 &\tr \bigl(X_N(\Omega)\bigr)^2
\nonumber \\
 &\quad= \prod_{j=1}^N \frac{1}{(d_j-1)^2}
 \sum_{\Lambda,\Lambda'\subset\{1,2,\ldots,N\}}
 (-1)^{\abs{\Lambda} + \abs{\Lambda'}} \tr \Bigl( \bigl( \rho_\Lambda \otimes \idty_{\Lambda^{\r
 C}}\bigr)\, \bigl(\rho_{\Lambda'} \otimes \idty_{\Lambda'^{\r C}}\bigr) \Bigr)
\nonumber \\
 &\quad= \prod_{j=1}^N \frac{1}{(d_j-1)^2}
 \sum_{\Lambda,\Lambda'\subset\{1,2,\ldots,N\}}
 (-1)^{\abs{\Lambda} + \abs{\Lambda'}}
 \biggl( \prod_{j\in(\Lambda\cup\Lambda')^{\r C}} d_j\biggr) \tr \rho_{\Lambda\cap\Lambda'}^2
\nonumber \\
 &\quad= \prod_{j=1}^N \frac{1}{(d_j-1)^2}
 \sum_{\Lambda\subset\{1,2,\ldots,N\}} \tr \rho_\Lambda^2
 \sum_{\Delta\subset\Lambda^{\r C}} (-1)^{\abs{\Delta}}
 \sum_{\Delta'\subset\Lambda^{\r C}\setminus\Delta} (-1)^{\abs{\Delta'}}
 \prod_{j\in(\Lambda^{\r C}\setminus\Delta)\setminus\Delta'} d_j
\nonumber \\
 &\quad=\prod_{j=1}^N \frac{1}{(d_j-1)^2}
 \sum_{\Lambda\subset\{1,2,\ldots,N\}} \tr \rho_\Lambda^2\,
 \prod_{j\in\Lambda^{\r C}} (d_j-2)
 \le \prod_{j=1}^N \frac{1}{d_j-1}.
\label{ineq}
\end{align}
In this computation $\Lambda^{\r C}$ denotes the complement of $\Lambda$ in
$\{1,2,\ldots,N\}$ and we used $\tr \bigl(\rho_\Lambda\bigr)^2 \le 1$.

Finally, we use the inequality
\begin{equation*}
 \r S_2(\rho) \le \r S_p(\rho) \le \r S(\rho),\quad 1\le p\le2
\end{equation*}
relating the $p$-Renyi entropies and the 2-Renyi entropy~\cite{ren} to obtain
\begin{equation*}
 \meo_p\bigl[\otimes_{j=1}^N \Gamma_{d_j}\bigr] \ge -\log\biggl(\prod_{j=1}^N
 \frac{1}{d_j-1} \biggr) = \sum_{j=1}^N \meo_p\bigl[\Gamma_{d_j}\bigr],
\end{equation*}
proving hereby~(\ref{add}).
\bigskip

\noindent
\textbf{Acknowledgements:}
R.A. is grateful for the hospitality during his stay at the K.U.Leuven and
acknowledges the financial support by the grant PBZ-MIN-008/P03/2003 of the
Polish Ministry of Scientific Research and Information Technology and EC
grant RESQ contract IST-2001-37559.
This work was also partially supported by F.W.O., Vlaanderen grant G.0109.01.
The authors also thank the referee for helpful suggestions concerning
the presentation of their result.


\begin{thebibliography}{99}

\bibitem{dat}
 N.~Datta, A.S.~Holevo and Y.M.~Suhov:
 A quantum channel with additive minimum output entropy,
 quant-ph/0403072
\bibitem{hol}
 A.S.~Holevo:
 Remarks on the classical capacity of quantum channel,
 quant-ph/0212025
\bibitem{kin}
 C.~King and M.B.~Ruskai:
 Comments on multiplicativity of maximal $p$-norms when $p=2$,
 quant-phys/0401026
\bibitem{mat}
 K.~Matsumoto T.~Shimono and A.~Winter:
 Remarks on additivity of the Holevo channel capacity and of the
 entanglement of formation,
 \emph{Commun.\ Math.\ Phys.}\textbf{246}, 427--442 (2004)
\bibitem{ren}
 A.~Renyi:
 \emph{Probability Theory},
 Akademiai Kiado, Budapest, North-Holland, Amsterdam (1970)
\bibitem{sho}
 P.~Shor:
 Equivalence of additivity questions in quantum information theory,
 quant-ph/0305035
\bibitem{wer}
 R.F.~Werner and A.S.~Holevo:
 Counterexample to an additivity conjecture for output purity of quantum
 channels,
 \emph{J.\ Math.\ Phys.\ }\textbf{43}, 4353 (2002)
 quant-ph/0203003
\end{thebibliography}
\end{document}